\documentclass[12pt]{article}%
\usepackage{amsmath}
\usepackage{amsfonts}
\usepackage{amssymb}
\usepackage{graphicx}%
\begin{document}

\begin{center}
{\Large \textbf{The Advantage of Rightmost Ordering for $\gamma_{5}$ in
Dimensional Regularization}}

\bigskip\bigskip

{\large Er-Cheng Tsai}\footnote{Email address: ectsai@ntu.edu.tw}

\bigskip

\textit{Physics Department, National Taiwan University, Taipei, Taiwan}

\bigskip

\hrulefill

\bigskip

\textbf{{Abstract} }
\end{center}

We propose a $\gamma_{5}$ scheme in dimensional regularization by analytically
continuing the dimension after all the $\gamma_{5}$ matrices have been moved
to the rightmost position. All Feynman amplitudes corresponding to diagrams
with no fermion loops regulated in this manner automatically satisfy the
Ward-Takahashi identities. This is in contrast to the scheme of Breitenlohner
and Maison, in which finite counter-terms are needed to restore gauge
invariance. This rightmost $\gamma_{5}$ scheme also has an advantage over the
naive dimensional regularization scheme which does not have a definitive
prescription consistent with gauge symmetry. Diagrams with fermion loops can
be handled by selecting a proper cut point on each fermion loop to play the
role of the point of the rightmost position.

\begin{center}
\hrulefill
\end{center}

\bigskip\bigskip

The difficulty of handling $\gamma_{5}$ in dimensional regularization
\cite{HV} stems from the inability of extending the anti-commutation
relationship%
\begin{equation}
\gamma_{5}\gamma_{\mu}+\gamma_{\mu}\gamma_{5}=0 \label{g5ant}%
\end{equation}
to space-time dimension $n\neq4$. If this relationship is forsaken under
dimensional regularization, a gauge theory involving chiral fermions is not
gauge invariant and the perturbative amplitudes calculated therewith may not
satisfy all the relevant Ward-Takahashi \cite{WTI}\ or BRST \cite{BRS} identities.

One way to overcome this difficulty is to use the Breitenlohner-Maison
\cite{BM} scheme. In this scheme, the $\gamma_{5}$ matrix anti-commutates with
$\gamma^{\mu}$ for $\mu\in\left\{  0,1,2,3\right\}  $\ but commutes with
$\gamma^{\mu}$ when $\mu$ is continued beyond the first four dimensions. This
BM scheme is not gauge invariant and finite counter-term renormalizations are
required to restore the validities of Ward-Takahashi identities
\cite{GB2,TLTM,MS,SANR}. This is seeking help from Dyson's scheme of
renormalization \cite{Dyson}. Such a mixed scheme adds to the complications of
calculations while reducing the conceptual simplicity of dimensional renormalization.

We shall present a simple method to deal with $\gamma_{5}$ which reserves the
validities of the Ward identities. This is done by exploiting the
anti-commutation relationship $\left(  \ref{g5ant}\right)  $ which allows us
to move all the $\gamma_{5}$ matrices to a proper position before continuing
it to $n\neq4$. For an open fermion line, this proper position is the
rightmost position. The proper position for a closed fermion loop will be
defined later.

There is a naive dimensional regularization (NDR) scheme which assumes that
$\gamma_{5}$ satisfies $\left(  \ref{g5ant}\right)  $ for all $\mu$ when
$n\neq4$. Since no such $\gamma_{5}$\ exists, this scheme is not without
fault. In particular, it is not capable of producing the triangular anomaly
term \cite{ABJ}. While regulated amplitudes satisfying Ward-Takahashi
identities have often been obtained in the past with the use of the NDR scheme
\cite{CFH,BDGKS,FER}, it is because all the $\gamma_{5}$ matrices have been
tacitly moved outside of divergent sub-diagrams in these calculations. This is
to say that the rightmost $\gamma_{5}$ scheme has been employed in actuality.

Let us introduce the notation $\underline{p^{\mu}}$ for the component of
$p^{\mu}$ vector in the first 4 dimensions and the notation $p_{\Delta}^{\mu}$
for the component in the remaining $n-4$ dimensions. $i.e.$, $p^{\mu
}=\underline{p}^{\mu}+p_{\Delta}^{\mu}$ with $p_{\Delta}^{\mu}=0$\ if $\mu
\in\left\{  0,1,2,3\right\}  $ and $\underline{p}^{\mu}=0$\ if $\mu
\notin\left\{  0,1,2,3\right\}  $. Likewise, the Dirac matrix $\gamma^{\mu}$
is decomposed as $\gamma^{\mu}=\underline{\gamma}^{\mu}+\gamma_{\Delta}^{\mu}$
with $\gamma_{\Delta}^{\mu}=0$\ when $\mu\in\left\{  0,1,2,3\right\}  $ and
$\underline{\gamma}^{\mu}=0$ when $\mu\notin\left\{  0,1,2,3\right\}  $. We
maintain the definition of $\gamma_{5}$ as its original one in the
four-dimensional space.
\begin{equation}
\gamma_{5}=i\gamma^{0}\gamma^{1}\gamma^{2}\gamma^{3} \label{g5def}%
\end{equation}
This definition of $\gamma_{5}$ satisfies $\gamma_{5}^{2}=1$ and the
anti-commutation relationship $\gamma_{5}\gamma_{\mu}+\gamma_{\mu}\gamma
_{5}=0$ for $\mu\in\left\{  0,1,2,3\right\}  $. But when $\mu$ is not in
$\left\{  0,1,2,3\right\}  $,
\begin{equation}
\gamma_{5}\gamma^{\mu}+\gamma^{\mu}\gamma_{5}=2\gamma_{\Delta}^{\mu}\gamma
_{5}, \label{g5ant2}%
\end{equation}
and $\gamma_{5}$ does not anti-commute with $\gamma^{\mu}$. In a
four-dimensional space, any matrix product
\[
\hat{M}=\gamma_{\omega_{1}}\gamma_{\omega_{2}}...\gamma_{\omega_{n}}\text{
with }\omega_{i}\in\left\{  0,1,2,3,5\right\}
\]
may be reduced, by anti-commuting $\gamma_{5}$ to the rightmost position, to
either the form of $\pm\gamma_{\mu_{1}}\gamma_{\mu_{2}}...\gamma_{\mu_{m}}$
with $\mu_{i}\in\left\{  0,1,2,3\right\}  $ if $\hat{M}$ contains even
$\gamma_{5}$ factors, or the form $\pm\gamma_{\nu_{1}}\gamma_{\nu_{2}%
}...\gamma_{\nu_{p}}\gamma_{5}$ with $\nu_{i}\in\left\{  0,1,2,3\right\}  $ if
the $\gamma_{5}$ count is odd. As the $\gamma_{\mu}$ matrix is analytically
continued and consistently defined when the component $\mu$ runs out of the
range $\left\{  0,1,2,3\right\}  $ under the dimensional regularization
scheme, the matrix product $\gamma_{\mu_{1}}\gamma_{\mu_{2}}...\gamma_{\mu
_{m}}$ is also unambiguously defined under dimensional regularization. We may
also analytically continue the product $\gamma_{\nu_{1}}\gamma_{\nu_{2}%
}...\gamma_{\nu_{p}}\gamma_{5} $ with one $\gamma_{5}$ on the right by
defining it to be the product of the analytically continued $\gamma_{\nu_{1}%
}\gamma_{\nu_{2}}...\gamma_{\nu_{p}}$ and the $\gamma_{5}$ defined in $\left(
\ref{g5def}\right)  $. Similarly, we may analytically continue the product
$\gamma_{\nu_{1}}\gamma_{\nu_{2}}...\gamma_{\nu_{i}}\gamma_{5}\gamma
_{\nu_{i+1}}...\gamma_{\nu_{p}}$ by defining it to be the analytically
continued $\gamma_{\nu_{1}}\gamma_{\nu_{2}}...\gamma_{\nu_{i}}$ times
$\gamma_{5}$ then times the analytically continued $\gamma_{\nu_{i+1}%
}...\gamma_{\nu_{p}}$.

In a $n=4$ dimensional space, a matrix product involving one $\gamma_{5}$ has
more than one equivalent expressions corresponding to different positionings
of the $\gamma_{5}$ matrix such as%
\[
\gamma_{\nu_{1}}\gamma_{\nu_{2}}...\gamma_{\nu_{p}}\gamma_{5}=\left(
-1\right)  ^{p-i}\gamma_{\nu_{1}}\gamma_{\nu_{2}}..\gamma_{\nu_{i}}\gamma
_{5}\gamma_{\nu_{i+1}}..\gamma_{\nu_{p}}%
\]
for $i=0,1,..p-1$. When $n\neq4$, the above equation does not always hold
because the anti-commutator $\left(  \ref{g5ant}\right)  $ becomes commutor
$\left(  \ref{g5ant2}\right)  $ and does not vanish. In particular, we have%
\[
\gamma_{\nu_{1}}\gamma_{\nu_{2}}...\gamma_{\nu_{p}}\gamma_{5}=-\gamma_{\nu
_{1}}\gamma_{\nu_{2}}...\gamma_{5}\gamma_{\nu_{p}}+2\gamma_{\nu_{1}}%
\gamma_{\nu_{2}}...\gamma_{\nu_{p-1}}\gamma_{5}\gamma_{\Delta,\nu_{p}}%
\]
Thus a matrix product that contains an odd number of $\gamma_{5}$ is not
unambiguously continued from its value at $n=4$.

Before analytic continuation is made, a $\gamma_{5}$-odd matrix product may
always be reduced to a matrix product with only one $\gamma_{5}$ factor. To
analytically continue such a matrix product, we need an extra information
which is specifying the location of the $\gamma_{5}$ factor within the matrix
product. We adopt the default continuation by anti-commuting the $\gamma_{5}$
matrix to the rightmost position before making the continuation.

For the QED theory, the identity%
\begin{equation}
\frac{1}{\not \ell +\not k  -m}\not k  \frac{1}{\not \ell -m}=\frac{1}{\ell
-m}-\frac{1}{\not \ell +\not k  -m} \label{qedwti0}%
\end{equation}
is the foundation that a Ward-Takahashi identity is built upon. For a gauge
theory involving $\gamma_{5}$, there is a basic identity similar to $\left(
\ref{qedwti0}\right)  $ for verifying Ward identities:%
\begin{equation}
\frac{1}{\not \ell +\not k  -m}\left(  \not k  -2m\right)  \gamma_{5}\frac
{1}{\not \ell -m}=\gamma_{5}\frac{1}{\not \ell -m}+\frac{1}{\not \ell +\not k
-m}\gamma_{5} \label{canc}%
\end{equation}
The above identity valid at $n=4$ is derived by decomposing the vertex factor
$\left(  \not k  -2m\right)  \gamma_{5}$ into $\left(  \not \ell +\not k
-m\right)  \gamma_{5}$ and $\gamma_{5}\left(  \not \ell -m\right)  $ that
annihilate respectively the propagators of the outgoing fermion with momentum
$\ell+k$ and the incoming fermion with momentum $\ell$. Positioning
$\gamma_{5}$ at the rightmost site, the above identity becomes%
\[
\frac{1}{\not \ell +\not k  -m}\left(  \not k  -2m\right)  \frac
{1}{-\not \ell -m}\gamma_{5}=\left(  \frac{1}{-\not \ell -m}+\frac
{1}{\not \ell +\not k  -m}\right)  \gamma_{5}%
\]

If we disregard the rightmost $\gamma_{5}$ on both sides of the above
identity, we obtain another identity that is valid at $n=4$. This new
identity, which is void of $\gamma_{5}$, may be analytically continued to hold
when $n\neq4$. We then multiply $\gamma_{5}$ on the right to every
analytically continued term of this $\gamma_{5}$-free identity to yield the
analytic continuation of the identity $\left(  \ref{canc}\right)  $.

As a side remark, we note that when we go to the dimension of $n\neq4$,
$\left(  \ref{canc}\right)  $ in the form presented above is not valid. This
is because $\gamma_{5}$ does not always anti-commute with $\gamma^{\mu}$ if
$n\neq4$. Instead, the identity needs to be modified by including an
additional vertex factor $2\not \ell _{\Delta}\gamma_{5}$, as shown below, if
it is to hold for $n\neq4$.%
\[
\frac{1}{\not \ell +\not k  -m}\left(  \not k  +2\not \ell _{\Delta
}-2m\right)  \gamma_{5}\frac{1}{\not \ell -m}=\gamma_{5}\frac{1}{\not \ell
-m}+\frac{1}{\not \ell +\not k  -m}\gamma_{5}%
\]
Adopting the rightmost $\gamma_{5}$ ordering avoids this difficulty, as the
validity of the identity in the form of rightmost $\gamma_{5}$ ordering no
longer depends on $\gamma_{5}$ anti-commuting with the $\gamma$ matrices.

For an amplitude corresponding to a diagram involving no fermion loops, we
shall move all $\gamma_{5}$ matrices to the rightmost position before we
continue analytically the dimension $n$. Subsequent application of dimensional
regularization gives us regulated amplitudes satisfying the Ward identities.

If a diagram has one or more fermion loops, the amplitude corresponding to
this diagram can be regulated in more than one ways. This is because there are
different ways to assign the starting position on a fermion loop. Once we have
chosen a starting point, we define the matrix product inside the trace by
rightmost $\gamma_{5}$ ordering before making the analytic continuation. In
general, continuations from different starting points give different values
for the trace when $n\neq4$.

An identity relating the traces of matrix products without $\gamma_{5}$ at
$n=4$ can always be analytically continued to hold when $n\neq4$. Therefore,
the portion of an amplitude in which the count of $\gamma_{5}$ on every loop
is even has no $\gamma_{5}$ difficulty \cite{GD}. But to calculate amplitudes
with an odd count of $\gamma_{5}$, we need an additional prescription. This is
because, as we have mentioned, the rightmost position on a fermion loop is not
defined a-priori.

We note that a fermion loop opens up and becomes a fermion line if we make a
cut at some point on the loop. We shall always choose as the cut point either
the beginning point or the endpoint of an internal fermion line on the loop.
An internal fermion line begins from a vertex and ends at another vertex. When
the cut point is chosen to be the endpoint of an internal fermion line, the
vertex factor will be assigned to appear as the beginning factor and stands at
the right end of the matrix product for the entire open fermion line. And when
the cut point is chosen to be the beginning point of an internal fermion line
that emits from a vertex, the matrix factor corresponding to that vertex will
be assigned to be the terminating factor and stands at the left end of the
matrix product for the entire open fermion line. With the cut point on a
fermion loop chosen and with the fermion loop turned into a fermion line, we
may apply the rule of rightmost ordering for $\gamma_{5}$.

Although we have multiple continuations for a matrix product or the trace of a
matrix product, they differ with one another either by terms that are
$O\left(  n-4\right)  $ or by terms containing at least a factor of
$\gamma_{\Delta}^{\mu}$. In the tree order and in the limit $n\rightarrow4$,
they are all restored to the same result because $\gamma_{\Delta}^{\mu}$ will
disappear when $n\rightarrow4$. For higher loop orders, $\gamma_{\Delta}^{\mu
}$ contribution may not be ignored in the limit $n\rightarrow4$. This is
because the factor $\gamma_{\Delta}^{\mu}\gamma_{\Delta}^{\nu}g_{\mu\nu
}=\left(  n-4\right)  $ becomes finite or even infinite in the limit
$n\rightarrow4$ if it is multiplied by a divergent integral which generates a
simple pole factor $\frac{1}{\left(  n-4\right)  }$, or terms of higher pole
orders. Thus $\gamma_{5}$ located within a divergent diagram or sub-diagram in
general yields different regulated amplitude from that given by rightmost
$\gamma_{5}$.

A $\gamma_{5}$ position will be called proper if it is not located within a
divergent 1PI sub-diagram such as a self-energy insertion or a vertex
correction. Likewise, for a fermion loop, a cut and the corresponding cut
point will be called proper if the cut is not made within a divergent
self-energy insertion or vertex correction sub-diagram.

The minimal subtraction prescription, which subtracts out the pole terms for
all possible forests of non-overlapping sub-diagrams \cite{ZIM,GTH,BM}, is a
convenient renormalization procedure. For a superficially convergent diagram
with an open fermion line or a closed fermion loop that has been cut open at a
proper cut point, positioning $\gamma_{5}$ at any other proper location also
gives the same amplitude as the default continuation with rightmost
$\gamma_{5}$ in the limit $n\rightarrow4$ if all the divergent sub-diagrams
have been renormalized. Therefore, the renormalized amplitudes for a
superficially convergent 1PI diagram obtained with different proper
$\gamma_{5}$ locations or proper cut points approach the same $n\rightarrow4$ limit.

But for a superficially divergent diagram, even with all proper sub-diagrams
renormalized, pole terms may still arise from overall integrations. If we
expand the overall amplitude in a Taylor series with respect to the external
momenta, the pole terms occur only in the first few terms in the series
because the degree of divergence from power counting for each term in the
Taylor series is progressively decreased by the power of the external momenta.
For example, the vertex correction function is logarithmically divergent and
only the term with all the external momenta set to zero may have pole terms if
all proper sub-diagrams have been renormalized. The overall subtraction of
pole terms does not remove the finite difference stemming from multiplying
these overall pole terms to the $\gamma_{\Delta}$ or $\left(  n-4\right)  $
difference even if we position $\gamma_{5}$ at two different proper locations.

If we rely on the Ward identities to determine these ambiguous finite terms as
is done in the BM scheme, we can choose whichever position or cut point for
$\gamma_{5}$ as long as it is a proper one. The renormalized amplitudes so
calculated are consistent with those obtained from the BM scheme. But this
method of finite counter-term renormalization is rather complicated and
difficult to implement in practical calculation. Fortunately, we are able to
show in \cite{ECSM} that there is a cut-point prescription which is capable of
regularizing amplitudes gauge invariantly under the rightmost $\gamma_{5}$
scheme for diagrams up to 2-loop order provided that the 1-loop triangular
anomaly is absent. This result is significant in that it greatly reduces the
complexity of calculations for amplitudes in the standard model. Furthermore,
by moving all the $\gamma_{5}$ to the rightmost position, the matrix product
in front of $\gamma_{5}$ is a fully $n$-dimensional covariant expression and,
in contrast to the non-covariant treatment of the $\gamma$ matrix indices in
the BM scheme, we are spared the chore of splitting the $n$ dimensional space
into $4$ and $\left(  n-4\right)  $ spaces in practical calculations. Thus the
rightmost $\gamma_{5}$ scheme is a much simpler scheme than the BM scheme for
calculating renormalized amplitudes in the standard model.

To summarize, the prescription of anti-commuting $\gamma_{5}$ to the rightmost
position before applying dimensional regularization ensures that all Ward
identities are regularized and remain valid for $n\neq4$. This means that an
amplitude corresponding to a diagram with no fermion loops can always be
renormalized with dimensional regularization and minimal subtractions. The
renormalized amplitudes so obtained satisfy the Ward identities and no
additional finite counter-term renormalization is needed in this rightmost
$\gamma_{5}$ scheme. Divergent diagrams with fermion loops are the only type
of diagrams that may be ambiguous with respect to the $\gamma_{5}$
positioning. Not incidentally, they are also the diagrams that may be plagued
by anomaly problem. Such diagrams can also be handled by the rightmost
$\gamma_{5}$ scheme and will be treated in another paper \cite{ECSM}.

\end{document}